\documentclass{emulateapj}
\usepackage{latexsym}
\usepackage{dcolumn}
\usepackage{bm}
\input{epsf}
\newcommand{\be}{\begin{equation}}
\newcommand{\ee}{\end{equation}}
\newcommand{\ba}{\begin{eqnarray}}
\newcommand{\ea}{\end{eqnarray}}
\begin{document}

\title{A new method of measuring the peculiar velocity power spectrum}
\author{Pengjie Zhang\altaffilmark{1}, Albert
Stebbins\altaffilmark{1}, Roman Juszkiewicz\altaffilmark{2},Hume Feldman\altaffilmark{3}}
\email{zhangpj@fnal.gov, stebbins@fnal.gov, roman@camk.edu.pl, feldman@ku.edu}
\altaffiltext{1}{NASA/Fermilab Astrophysics Center,
Fermi National Accelerator Laboratory, Batavia, IL 60510-0500}
\altaffiltext{2}{Copernicus Astronomical Center, Warsaw, Poland}
\altaffiltext{3}{Dept. of Physics \& Astronomy, Univ. of Kansas, Lawrence, KS
66045, USA}
\begin{abstract}
We show that by directly correlating the cluster kinetic Sunyaev
Zeldovich (KSZ)  flux, the cluster
peculiar velocity power spectrum can be measured to $\sim 10\%$
accuracy by future large sky coverage KSZ surveys. This method is
almost free of systemics entangled in the usual
velocity inversion method. The direct correlation brings extra
information of density and velocity clustering. We  utilize these
information  to construct two indicators of  the Hubble constant  and
comoving angular distance and propose a   novel method to  
constrain cosmology.   
\end{abstract}
\section{Introduction}
The large scale peculiar velocity field, arising from primordial
density perturbation, is a fair tracer of the large scale structure of
the universe  
and  is of great importance to constrain cosmology and the nature of
gravity.  But the huge Hubble flow contamination makes its measurement
extremely difficult, though several important progresses have been
made (e.g. \citet{Davis83,Feldman03}).  The cluster kinetic Sunyaev
Zeldovich (KSZ) effect, which  directly measures the cluster peculiar momentum
with respect to the CMB,  opens a new window for the measurement of
the peculiar velocity $v_p$
\citep{Haehnelt96,Kashlinsky00,Aghanim01,Atrio-Barandela04,Holder04}. 
But the direct inversion of $v_p$ requires extra measurements of the
cluster Thomson optical depth $\tau$ and cluster temperature.
Contaminations in the thermal Sunyaev Zeldovich (TSZ) effect and KSZ
measurements and 
inappropriate  
assumptions involved in this process put a systematic limit of $\sim 200$
km/s to the  inferred $v_p$
\citep{Knox03,Aghanim04,Diaferio04}.  

A natural solution is to avoid the $v_p$ inversion, but use the KSZ flux, the  
direct observable.  Contaminations to
the cluster KSZ   flux have different  clustering properties and can
be applied to disentangle KSZ 
signal from contaminations. In the direct correlation estimator of the
KSZ flux, many contaminations automatically vanish and most remaining
contaminations can be  subtracted. We will show that at $z\ga 0.3$,
the systematics become 
sub-dominant and the statistical errors, at $\sim 10\%$ level for
South Pole telescope (SPT\footnote{http://astro.uchicago.edu/spt}),
dominates. We further derive a novel method, which measures
the Hubble constant and comoving distance and thus least model
dependent,  to constrain cosmology. Throughout this paper, we assume
$\Omega_m=0.3$, $\Omega_{\Lambda}=1-\Omega_m$, 
$\sigma_8=0.9$ and adopt BBKS transfer function \citep{Bardeen86}. 

\section{The flux power spectrum} 
The KSZ cluster surveys directly measure the sum of cluster KSZ flux
$S_{\rm KSZ}$ and
various contaminations, such as intracluster gas internal
flow,  radio and IR point sources, primary CMB,
cosmic infrared background (CIB), etc. The signal is 
\be
S_{\rm KSZ}
=\frac{\partial B_{T}(\nu)}{\partial T} T_{\rm CMB} \langle \tau \rangle v_p
\Delta\Omega_A =S_{100}v_{100}\ .
\ee
Here, $\langle \tau\rangle$ is the $\tau$  averaged over the solid
angle $\Delta \Omega_A$, $S_{100}$ is the KSZ flux assuming $v_p=100
{\rm km}/s$ and $v_{100}\equiv v_p/100 {\rm km}/s$.   Since at
$\nu\sim 217$ Ghz, the non relativistic TSZ effect, which is one of the major contaminations of the KSZ,
vanishes, throughout this paper, we focus on this frequency. 
 At $\nu\sim 217 $Ghz,
$\partial B/\partial T=540 \ {\rm Jy}\ {\rm sr}^{-1}\ {\rm
uK}^{-1}$.

Optical follow up of KSZ surveys such as dark energy survey will
measure cluster redshift $z$ with uncertainty  $\la
0.005$.\footnote{The photo-$z$ of 
each galaxies has dispersion $\sim 
0.05$. Clusters have $\ga 100$ galaxies and thus the determined $z$
dispersion is $\la 0.005$.} The $z$ information allows the
measurement of 3D correlation 
\ba
\label{eqn:xis}
\xi_{S}(r)&\equiv& \langle (S_i-\bar{I}\Delta
\Omega_i)(S_j-\bar{I}\Delta \Omega_j)\rangle\\
&=&\langle S_{{\rm KSZ},i}S_{{\rm KSZ},j} \rangle+\cdots\ . \nonumber
\ea
Here, $\bar{I}=\sum S_i/\sum \Delta \Omega_i$ and $\Delta \Omega$ is
the solid angle of each cluster. 
  In
the correlation estimator,  many contaminations such as internal flow
and instrumental noise vanish. The 
majority of remaining 
systematics can be subtracted directly and the residual
systematics is generally  (much) less than the statistical errors (\S \ref{sec:error}).

Since $\langle S_{\rm KSZ}S_{\rm KSZ} \rangle$ is cluster number
weighted and since $v_{100}$ is correlated at large scale, the
corresponding KSZ power spectrum $\Delta^2_{\rm KSZ}$ has
contributions from both the velocity power spectrum $\Delta^2_{v_{100}}$
and the power spectrum $\Delta^2_{\delta v\delta v}$ of $\langle
\delta_1v_1\delta_2v_2\rangle$, where $\delta$ is the matter
overdensity. We then have
\begin{eqnarray}
\label{eqn:ksz}
\Delta^2_{\rm KSZ}(k)&=&\left(\int_{m_{\rm low}} S_{100}(m) \frac{dn}{dm}dm 
\right)^2
\Delta^2_{v_{100}}(k) \\
&+&\left(\int_{m_{\rm low}}S_{100}(m) b_{n}(m) \frac{dn}{dm}dm 
\right)^2 \Delta^2_{\delta_1 v_{1}
\delta_2v_{2}}\ . \nonumber
\end{eqnarray}
Here, $dn/dm$ is the cluster  mass function. Throughout this paper,
we assume a unity velocity bias and 
cluster number   density bias $\langle b_{n}\rangle=3$.
$\Delta^2_{v_{100}}$ is given 
by the linear theory prediction  
\begin{equation}
\label{eqn:v}
\Delta^2_{v_{100}}(k)=\frac{1}{3}\beta^2 E^2(a)a^2
\frac{\Delta^2_{\rm DM}(k,z)}{k^2} \ .
\end{equation}
Here, $k$ is in unit of $h/$Mpc and $\beta\equiv (a/D) dD/da$ where
$D$ is the linear density growth factor. $E(a)$ is the evolution of
Hubble constant normalized to $E(a=1)=1$. $\Delta^2_{\rm DM}$ is the
dark matter power spectrum (variance). We have
assumed that two lines of sight are parallel to each other. For less
than $10^{\circ}$ angular separation, the accuracy of this assumption is
better than $\sim 1\%$.  
Because of the large bias
$\langle b_{n}\rangle$, the  $\langle \delta_1 v_{1}
\delta_2v_{2}\rangle$ term becomes dominant even in the
linear regime at $k\ga 0.06h/$Mpc.   Because the nonlinear scale at
$z\sim 1$ is $k\simeq0.5h/$Mpc, we can still  using the perturbation theory
to predict
\begin{eqnarray}
\label{eqn:dvdv}
\Delta^2_{\delta_1 v_{1}\delta_2v_{2}}({\bf
k})&=&\frac{1}{6} \beta^2E^2(a)a^2 \frac{k^3}{2\pi^2} \times 
\\ &&\int 
P_{\rm DM}(|{\bf k}-{\bf 
k}_2|)\frac{k^2}{|{\bf k}-{\bf
k}_2|^2}\frac{\Delta^2_{\rm DM}(k_2)}{k_2^2}d^3k_2 \nonumber \\
&\rightarrow&\frac{1}{2}\Delta^2_{\rm DM}(k,z) \sigma^2_v(z)\ \ \ {\rm
when}\ k\ga 0.15 h/{\rm Mpc} \nonumber
\end{eqnarray}

Here, $\sigma_{v_{100}}^2$ is the one dimensional velocity 
dispersion in unit of $(100 {\rm km}/s)^2$. $v_p$ is determined by large
scale gravitational potential, so it decouples from  small scale density
fluctuation. Thus the last expression of Eq. \ref{eqn:dvdv} holds even
in the nonlinear regime,   if substituting the two quantities with the
corresponding nonlinear ones.   

\begin{figure}
\epsfxsize=9cm
\epsffile{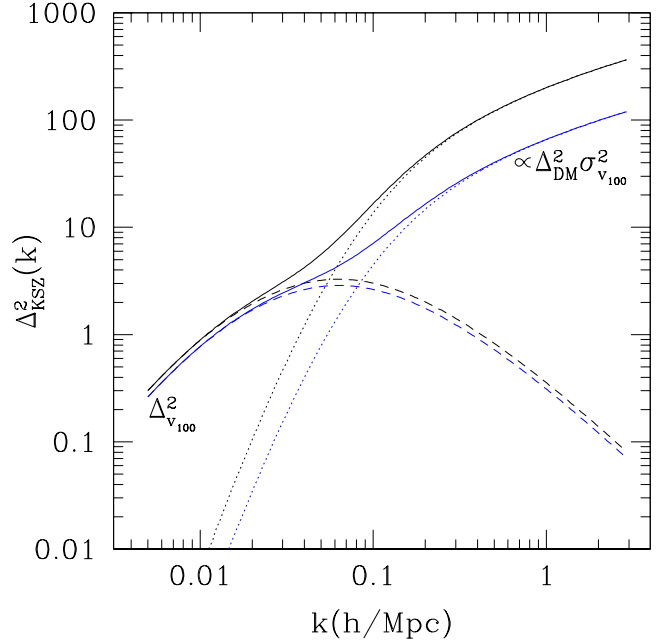}
\caption{The normalized KSZ cluster flux power spectra at $z=0$ (black
line) and $z=1$ (blue) respectively. The dash lines are the
contributions from the $vv$ term and the dot lines are the
contributions from the $\delta v\delta v$ term. The solid line is the
sum of them. \label{fig:v}}
\end{figure}

The KSZ signal is amplified in the correlation, with respect to many
contaminations due to three reasons. (1) Because of the $k^2$ denominator,
$\Delta^2_v$ peaks at $k\sim 0.05h/$Mpc, while $\Delta^2_{\rm DM}$, which many contaminations follow, keeps
decreasing toward large scales.  (2) The extra
$\Delta^2_{\delta_1v_1\delta_2v_2}$ term causes $\Delta^2_{\rm KSZ}$ to
keep increasing toward large $k$. (3) $\Delta^2_{\rm KSZ}$ has much weaker
$z$ dependence at $k\la 0.03h/$Mpc (Fig. \ref{fig:v}), comparing to
$\Delta^2_{\rm  DM}$.

We have not considered the effect of redshift distortion in the above
calculation. One can show that the redshift distortion is negligible
at $k\la 0.1 
h/$Mpc in the $\Delta^2_v$ part. Its effect to $\Delta^2_{\delta
v\delta v}$ can be dealt with by usual methods
(e.g. \citet{Kaiser87}). While the overall effect 
of the redshift distortion is to increase the signal at $k\la 0.4
h/$Mpc, it reduces the signal at $k\ga 0.4 h/$Mpc and
thus makes the measurement of $\Delta^2_{\rm KSZ}$ more
difficult. Since we are mainly interested in the linear regime, for
simplicity, we  disregard this effect. 

\section{Noise estimation}
\label{sec:error}
We target at SPT to estimate the accuracy of $\Delta^2_{\rm KSZ}$
measurement. We assume that SPT will cover the $217$ Ghz range and 
$4400\  {\rm deg}^2$ with arc-minute resolution.  We discuss the three
dominant error sources, 
diffuse foregrounds and backgrounds (\S 
\ref{subsec:fb}), sources associated with clusters (\S
\ref{subsec:clst}) and cosmic variance and shot noise (\S
\ref{subsec:stat}). We assume redshift bin size $\Delta z=0.2$ and $k$
bin size $\Delta k=0.5k$.  

\subsection{Diffuse foregrounds and backgrounds}
\label{subsec:fb}
The KSZ signal has typical $\Delta T_{\rm KSZ}
\simeq 20 \mu K \langle\tau\rangle /0.01$.  Primary CMB, which is
indistinguishable from the KSZ  
signal in frequency space,  has intrinsic temperature
fluctuation $\Delta T \sim 100 \mu$K. The CIB has mean  $T\sim 20 \mu$K, if scaled to $217$ Ghz
\citep{Fixsen98}.  
A cluster KSZ filter 
can be applied to  filter away the mean backgrounds while keeping the KSZ
signal. This filter must strongly match the cluster
KSZ profile while having  zero integrated area. Since the angular size
of clusters are generally several arc-minute, such filter naturally
peaks at multipole $l$ around several thousands and thus naturally
filters away the dominant CMB signal, which concentrates at
$l\la 1500$.  For such filters, at $\sim 217$ Ghz, the galactic
synchrotron,  free-free foregrounds,  the 
galactic dust emission, the radio background, the TSZ background are
all negligible due to their frequency or scale dependence
(see, e.g. \citet{Wright98, Bennett03}). So we only discuss the 
contaminations of  the primary CMB, CIB and background  KSZ. 


The optimal filter can be constructed from the intracluster gas
profile  inferred from the TSZ survey.  For simplicity, 
we choose the electron density profile as 
$n_e(r)\propto (1+r^2/r^2_c)^{-1}$ and a compensate Gaussian filter 
$W(l)=6(l/l_f)^2\exp(-(l/l_f)^2)$. 
For these particular choices, filtered KSZ temperature, 
$\tilde{\Delta} T_{\rm KSZ}$, peaks at $l_f\sim 1.1/\theta_c\sim 3800
(1^{'}/\theta_c)$, where $\theta_c$ is the angular core radius, and
the peak value is $\tilde{\Delta} T_{\rm KSZ}\simeq \Delta T_{\rm KSZ}\simeq 
9 v_{100} [\langle\tau\rangle/0.01]\mu K $. For simplicity, we
adopt this $l_f$ and $\theta_c=0.4 {\rm Mpc/h}/\chi(z)$ to estimate
the noise. Here, $\chi(z)$ is the comoving angular distance.   

The correlations of filtered
backgrounds (with zero mean flux), originating from both their intrinsic
correlations and cluster clustering, are  $
\tilde{\xi}_{b}(r)\sim [\int_{m_{\rm low}} \Delta \Omega_A
dn]^2[1+\langle b_{n}\rangle^2 \xi_{\rm DM}(r)]
\tilde{w}_{b}(\theta\sim r/\chi)$, where $\tilde{w}_{b}(\theta)$ is the
corresponding (filtered) background
angular correlation function. 

 The CMB contamination is sub-tractable and does not cause systematic error. The same cluster survey
measures both $\int \Delta 
\Omega_A dn$ and $\langle b_n\rangle^2\xi_{\rm DM}$, while the CMB
$C_l$ has been precisely measured by WMAP and 
can be precisely predicted by CMBFAST \citep{CMBFAST}. Thus the CMB
contamination  
$\tilde{\xi}_{\rm CMB}$  
can be straightforwardly predicted and subtracted from the correlation
estimator.  Thus only the  statistical errors caused by the CMB
intrinsic fluctuation over cluster regions remains. 

\begin{figure}
\epsfxsize=9cm
\epsffile{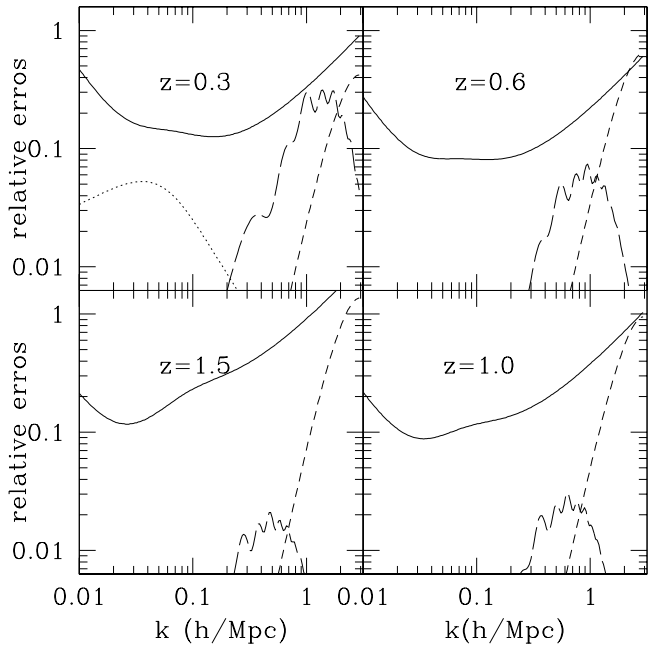}
\caption{The dominant errors of the cluster KSZ
power spectrum measurement. The solid lines are the errors caused by
the cosmic variance and shot noise. The short dash lines are the
systematic errors caused by the CIB and KSZ background. The long dash
lines are the statistical errors caused by the primary CMB. The dot
lines (which are  visible only in the upper left panel) are the
statistical errors caused by sources associated clusters. Systematic
errors are sub-dominant at $k\la 1h/$Mpc.   \label{fig:error}}
\end{figure}

The CIB and KSZ contaminations are in principle sub-tractable, too. But
since both the amplitude and shape of CIB and KSZ power spectra are highly
uncertain, we do not attempt to subtract their contribution from the
correlation estimator. The CIB power spectrum is $C_l^{\rm
CIB}l^2/(2\pi)\simeq (4 \mu {\rm K})^2 (l/10^3)^{0.7}$ (see,
e.g. \citet{Zhang03}) and the  kinetic SZ power spectrum is $C_l^{\rm
back-KSZ}l^2/(2\pi)\simeq (2.7 \mu {\rm K})^2$ \citep{Zhang04}. The
upper limit of the fractional systematic error they cause is  
\begin{eqnarray}
\eta&\sim& \frac{
\left[\sum_{b={\rm CIB,KSZ}}C^{b}_l\frac{l^2}{2\pi}W^2(l,l_f)\right]_{l\sim k\chi}}{[9  (\langle\tau\rangle/0.01)\mu K]^2}\times \nonumber \\
&&\frac{1+\langle b_{n}\rangle^2 \Delta^2_{\rm
DM}(k)}{\Delta^2_{v_{100}}(k)+\langle
b_{n}\rangle^2\Delta^2_{\delta_1v_1\delta_2v_2}}\ ,
\end{eqnarray}
which is at most several percent at $k\la 1h/$Mpc (Fig. \ref{fig:error}).


\subsection{Sources associated with clusters}
\label{subsec:clst}
The map filter does not filter away the contaminations
from sources associated with clusters.  At $217$Ghz, the
non-relativistic TSZ vanishes. But the relativistic
correction of cluster TSZ effect shifts the cross over point slightly to higher
frequency and thus in principle introduces a residual thermal SZ
signal in $\sim 217$Ghz band. We assume an effective $\Delta T_{\rm
TSZ}\sim 1\mu $K, or, $\sim 1\%$ of the TSZ at Rayleigh-Jeans
regime.  This gives a flux
$S_{\rm res-SZ}\sim 540 \ {\rm Jy}\ {\rm sr}^{-1}$.  The flux of radio
sources and IR sources associated with cluster is $\sim 10^3{\rm
Jy}{\rm sr}^{-1}$ at $217$ Ghz (see, e.g. \citet{Aghanim04}).  By
multi-frequency 
information and  resolved source subtraction, one is
likely able to subtract much of these contaminations. Rather
conservatively, we assume that, at $\nu\sim 217 $Ghz, the total 
flux contributed by those sources  associated with 
clusters is less than $\sim 5\times 
10^3 {\rm Jy}{\rm sr}^{-1}$ at $z=0.5$ and scale it to other redshifts
assuming no intrinsic luminosity evolution.

The mean flux of these sources is subtracted in our estimator
(Eq. \ref{eqn:xis}). Since the cluster thermal energy, IR and radio
flux  should be  mainly  determined by local processes,  one can omit
the possible large scale  correlation of these  quantities. Thus these
sources do not cause systematic errors. But since $\langle \delta S_{\rm
clst}^2\rangle\neq 0$, they do contribute to statistical errors.  The
fractional error they cause is
\ba
\eta &\sim & \frac{\langle \delta S_{\rm
clst}^2\rangle}{S^2_{100}} \frac{\langle 
b_{n}\rangle^2 \Delta^2_{\rm
DM} 2\pi[V k^2\Delta k]^{-1/2}}{\Delta^2_{v_{100}}+\langle
b_n\rangle^2\Delta^2_{\delta_1v_1\delta_2v_2}}\nonumber \\
&\la&  5\times 10^{-3}\frac{\langle \delta S_{\rm
clst}^2\rangle}{S^2_{100}} z^{-1} \sqrt{\frac{\Delta z}{0.2} \frac{\Delta k/k}{0.5}} 
\ea 
Here, $V$ is the survey volume in the adopted redshift bin. It is reasonable to
assume that  $\langle \delta S^2_{\rm clst}\rangle^{1/2}\la \langle
S_{\rm clst}\rangle$. Thus  the error caused by
the sources  associated with clusters  is negligible at almost all
scales and redshifts (Fig. \ref{fig:error})


\subsection{Cosmic variance and shot noises}
\label{subsec:stat}
The signal intrinsic cosmic variance  dominates at large
scales. The amount of cluster is very limited, 
so the shot noise is large, even in the linear scales. The
signal, the instrumental noise, the rms  
flux fluctuations of any contaminations projected onto or associated with
clusters, such as primary CMB, IR and radio sources, all contribute
to the shot noise, whose power spectrum is 
\be
\tilde{\Delta}^2_{\rm short}=\frac{k^3}{\bar{n}
2\pi^2}\left(\tilde{\sigma}^2_{\rm
CMB}+\tilde{\sigma}^2_{\rm CIB}+\tilde{\sigma}^2_{\rm 
KSZ}+\tilde{\sigma}^2_{\rm clst} +\cdots \right). 
\ee
Here, $\bar{n}$ is the mean  number density of observed clusters, which
can be calculated given the halo mass function, the survey
specification and the gas model. For simplicity,
we assume $\bar{n}(z)=3\times 10^{-5}/(1+z)^3 (h/{\rm Mpc})^3$. We only
consider the listed 4 dominant sources, where $\tilde{\sigma}_{\rm
CMB}\sim \tilde{\sigma}_{\rm CIB} \sim \tilde{\sigma}_{\rm KSZ} \sim
20 \mu K$ and $\tilde{\sigma}_{\rm  
clst}\sim 5\mu K$.
The beamed and filtered SPT instrumental noise has rms $\sim 1\mu$K
and is thus negligible. For SPT, the  error caused by the cosmic 
variance and shot noise  ($\sim 10\%$) dominates over all other errors. 
The systematic errors are (almost) always sub-dominant.   In this sense,
our method is  optimal to measure the cluster peculiar velocity. For a
future all sky  survey, total error can be reduced to several percent level.

\begin{figure}
\epsfxsize=9cm
\epsffile{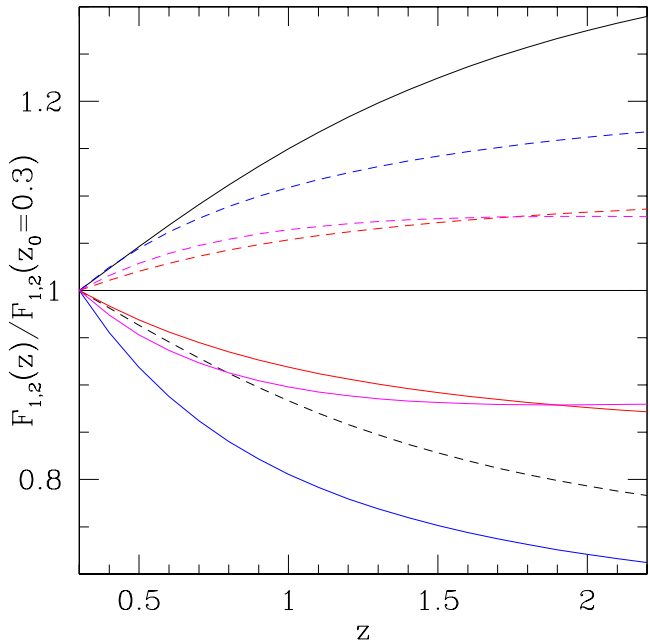}
\caption{The time variation of $F_{1}(k=0.015h/$Mpc (solid lines) and
$F_2(k=0.2h/{\rm Mpc},k_o=0.015h/{\rm Mpc}$
(dash lines)   caused by wrong assumed cosmological parameters. Our
fiducial model (black straight line) has $\Omega_m=0.3$ and
$\Omega_{\Lambda}=0.7$. Blue, 
magenta, red and black curves correspond to
$(\Omega_m,\Omega_{\Lambda})=(1.0,0.0), (0.3,0.0), (0.5,0.5)$ and
$(0.1,0.9)$, respectively.  \label{fig:r}}
\end{figure}
\section{Constraining cosmology}
There are several ways to constrain cosmology from the KSZ
observations, such as directly comparing the theoretical prediction of
Eq. \ref{eqn:ksz} with observations. Extra information contained in
the correlation allows us to develop a least model dependent method. 
Defining $\eta(k)\equiv\Delta^2_{\rm KSZ}(k)/\Delta^2_n(k)$, utilizing
the asymptotic behavior that $\eta(k\rightarrow 0)\rightarrow \langle
S_{100}\rangle^2 \beta^2E^2a^2/\langle b_n\rangle^2 3k^2$ and
$\eta(k\rightarrow \infty)\rightarrow \langle
S_{100}\rangle^2\sigma_v^2$, we obtain two observables,
$F_1(k)=\Delta^2_{\rm KSZ}(k)/\eta(k\rightarrow
\infty)=\Delta^2_v(k)/\sigma^2_v$ in the regime where $k\la 0.03h/{\rm
Mpc}$,  and $F_2(k,k_o)=\Delta^2_n(k)/[\eta(k\rightarrow
\infty)/\eta(k_o\rightarrow  0)]=\Delta^2_{\rm DM}(k)/[3\sigma^2_v
k_o^2/\beta^2E^2a^2]$ in the  linear regime where $k\la
0.3h/$Mpc. $F_{1,2}$ should be $z$ independent. 
Redshift distortion  and possible velocity bias do not change this
characteristic behavior. But  if the cosmology one choose is different
from the real cosmology, $k_w/k_r$, the ratio of 
determined wave-vector $k_w$ with respect to the real $k_r$, will
evolve with $z$. Because of the $k$ dependence of $F_1(k)$ and
$F_2(k)$, a wrong  cosmology will cause the determined $F_1$ and $F_2$
to vary with $z$. This behavior can be applied to constrain
cosmology.  One can show that
$k_{\parallel,w}/k_{\parallel,r}=E_w(z)/E_r(z)=g_{\parallel}(z)$
and
$k_{\perp,w}/k_{\perp,r}=\chi_r(z)/\chi_w(z)=g_{\perp}(z)$.
Any measured quantity
$\Psi(k)$ are then the corresponding quantity averaged
over all configurations, namely, $  
\langle \Psi(k,u)\rangle \equiv \int_{-1}^1
\Psi\left[k\sqrt{u^2/g_{\parallel}^2+(1-u^2)/g^2_{\perp}}\right
] du/2$.  We obtain
\begin{eqnarray}
\frac{F_{1,w}(z)}{F_{1,w}(z_0)}|_k& =&\frac{\langle
\Delta^2_v(k,u)\rangle}{\Delta^2_v(k)} \\
\frac{F_{2,w}(z)}{F_{2,w}(z_0)}|_{k,k_o} &=&\frac{\langle \Delta^2_{\rm
DM}(k,u)\rangle}{\Delta^2_{\rm DM}(k)}\frac{\Delta^2_{\rm
DM}(k_o)}{\langle \Delta^2_{\rm DM}(k_o,u)\rangle}\frac{\langle
\Delta^2_v(k_o,u)\rangle}{\Delta^2_v(k_o)} \ .\nonumber 
\end{eqnarray}
Fig. \ref{fig:r} shows the $z$ dependence of $F_{1,2}(z)$ for various assumed
cosmologies. This variation can reach $\sim 20\%$ and thus should be
detected by SPT. A future all sky survey would measure
this variation to better than $\sim 5\%$  accuracy and should provide a
tight and 
independent cosmology check.   The nonlinearity correction, the
parallel line of sight approximation, etc. may cause non-negligible
systematics. But these effects are straightforward to predict and
correct, so we postpone the discussion of these effects in this paper.

\section{Conclusion}
We present a new method to measure the peculiar velocity power
spectrum $\Delta^2_v$ by directly
correlating the KSZ flux.  The cluster peculiar velocity signal is
amplified in the direct 
correlation. Many systematics involved in the
usual $v_p$ inversion disappear and the majority of remaining
systematics can be easily subtracted. The correlation method is
optimal to measure $\Delta^2_v$, in the sense that statistical error,
dominates over 
systematics at almost 
all $k$ and $z$ range. We estimate that SPT can measure $\Delta^2_v$
to $\sim 10\%$ accuracy and a future all sky survey can improve this
measurement by a factor of several.  The KSZ flux correlation contains extra
information on the velocity dispersion. This extra information
helps to construct two independent observables  as indicators of
Hubble constant and comoving  angular distance, with only minimal amount of
assumptions. These observables should put  independent constraints on 
cosmology. 

{\it Acknowledgments}: P.J. Zhang and A. Stebbins are  supported by
the DOE and the NASA  grant NAG 5-10842 at Fermilab.

\end{document}